# Single-layer MoS$_2$ as efficient photocatalyst

Yunguo Li,*[a,c] Yan-ling Li,*[b,c] Carlos Moyses Araujo,[c] Wei Luo[c] and Rajeev Ahuja[a,c]

**Abstract**

**The potential application of the single-layer MoS$_2$ as photocatalyst was revealed based on first-principles calculations. It is found that the pristine single-layer MoS$_2$ is a good candidate for hydrogen production, and its catalysing ability can be tuned by the applied mechanical strain. Furthermore, the p-type doping could enable the single layer as an efficient self-consistent photocatalyst.**

## Introduction

The single-layer MoS$_2$ is an insulator with a direct bandgap of about 1.75-1.9 eV at the "K" point in Brillouin zone[1-8]. After the exfoliation from bulk[1], the 2D MoS$_2$ has been attracting extensive studies[7,9-11]. Recently, it has been revealed to exhibit exotic electronic properties that have illuminated many promising applications in valley electronics[7,9-11], transistor fabrications[12], optoelectronic devices[13] and high-end flexible electronics[14]. Besides, as mentioned by many researchers, it may also have important applications in photovoltaic applications in light of the tuneable band structures[1,3,5,6,12,14].

The bulk MoS$_2$ itself is not a good candidate for photocatalysts. However, the nanoscale MoS$_2$, like nanoribbon and cluster, exhibit unusual physical and chemical properties due to quantum confinement[13,15-18], and they possess photo catalyzing abilities[13,19,20]. It is therefore worthwhile to explore the photocatalytic property of the single-layer MoS$_2$. Besides, the single-layer MoS$_2$ possesses excellent charge carrier mobility, as good as the carbon nanotube[21]. Furthermore, the 2D sheet has a superior surface-to-volume ratio that could be an advantage for catalysing material. Therefore, we have explored in this work the potential of the single-layer MoS$_2$ as photocatalyst by means of calculations based on the density functional theory.

It is well known that the applied mechanical strain is a powerful and practical method to tune the electronic structures of materials[5,6]. Since the strain might exist in the fabrication process and during the forthputting of the single-layer MoS$_2$, we also have calculated its band structure and band edge energy levels with respect to the variant strain along zigzag direction, armchair direction as well as z-axis direction. Although Tianshu Li[5] and Emilio Scalise[6] have calculated the strain effects separately on the band structure along zigzag and armchair directions and given exciting results, we still believe it is necessary to do a comprehensive study on the strain effects both in in-plane directions and out-of-plane direction. Especially, the $p_z$ orbital often plays important role in 2D materials[22,23], and the single-layer MoS$_2$ actually contains three atomic layers with the Mo layer sandwiched by two S layers. So, the lattice change along z-direction is therefore also very important. Besides, the strain in z-axis direction is possible and much easier to be introduced during the preparation and characterization processes[1,14,24].

On the other hand, intrinsic and extrinsic doping on semiconductors are common strategies for the modification of electronic structures. The doping process is widely used in semiconductor industry, which has made great success in Si process, transistors, solar cells as well as photocatalysts. Doping with variant concentrations of dopants provides a powerful technique for people to manipulate the electronic structures of materials. The doping in semiconductors can not only modulate the bandgap, energy level, but also bring in modifications to the compositions of band edges. The donors or acceptors will induce charge redistribution to the doped system, so as to yield sites with totally different potentials, and thus affect the photocatalytic reaction. It is reasonable to expect that the photocatalytic activity can be tuned and improved by utilizing appropriate doping. So, we also have explored the p-type doping effects on the single-layer MoS$_2$ in search for the improvement of its catalysing ability.

## Computational methods

Our calculations were based on the first-principles density functional theory using projector augmented wave (PAW) method[25,26], which is implemented in the *Vienna Ab initio Simulation Package* (VASP)[27-29]. The generalized gradient approximation proposed by Perdew, Burke and Ernzrhof (GGA-PBE)[31] was used for the exchange and correlation functional. In addition, the hybridized (HSE06)[31,32] functional was also used for comparison to reveal the band nature of the single-layer MoS$_2$. A plane wave cut-off energy of 450 eV was found to be enough to reach convergence and used in all the calculations. The primitive cell of single-layer MoS$_2$ containing one Mo atom and two S atoms was used under periodic boundary conditions. A vacuum space of 16 Å adopted along z-axis was proved to be able to decouple possible periodic interactions. The valance electrons *3s3p* for S, *4p5s4d* for Mo and *3s3p* for P were included in the PAW pseudopotentials. The Brillouin zone was sampled using a Gamma centred 13×13×1 k-points mesh with Gaussian smearing method in structure relaxations, while a finer grid of 25×25×1 k-points was employed to conduct the following simulations. The used density of k-mesh was also tested to be enough to reach energy convergence.

In the context of photocatalytic reactions, the crucial step is the electron transfer between the photocatalyst and water redox species. Electrons can only be transferred between the states in the photocatalyst and the water redox species at approximately the same energy level. The standard redox potentials for water redox species (energy level of water redox species undergoing an electron transfer) are usually approximated with respect to the normal hydrogen electrode (NHE), which is frequently used in geochemical and electrochemical studies. Redox potentials contain the oxidation potential and the reduction potential. The oxidation potential and reduction potential at 25° is -5.67 eV and -4.44 eV, respectively, with respect to the absolute vacuum level[33], and the later is usually taken as zero by convention.

While the relevant energy levels for photocatalyst are the valence band maximum (VBM) and the conduction band minimum (CBM). The relative energetics of VBM and CBM vs. redox potentials is the important property of photocatalyst, which determines the feasibility of water splitting. Moreover, for ideal photocatalysts, the CBM should be located at a more negative potential than the proton reduction potential ($H^+/H_2$), while the VBM must be positioned more positively than the oxygen reduction potential ($O_2/H_2O$). To compare redox potentials of water with the band edge potentials of the photocatalyst, these potentials should be related to a common reference, for which we have used the absolute vacuum level. Here we have used a standard approach to refer the band edges to the vacuum level where the top of the valence band is assumed to be numerically equal to the materials work function. The latter is then calculated from the equation[34]:

$$\varphi = V_{(\infty)} - E_F \quad (1)$$

where $V_\infty$ is the electrostatic potential in a vacuum region, of a slab geometry calculation, far from the neutral surface and $E_F$ is the Fermi level of the neutral surface system. The reliability of this method has been verified by the weak temperature and pressure dependence of band edge potentials in the range of interest. Hence, in this way, we have succeeded in comparing the calculated band edge levels with experimental redox potentials in the same scale.

In view of the solar energy conversion efficiency, the high absorption rate of visible light is essential for photocatalysts. The optical properties of materials can be completely described by the dielectric function: $\varepsilon(\omega) = \varepsilon_1(\omega) + i\varepsilon_2(\omega)$, at all photon energies. The dielectric function was calculated in the momentum representation, by acquiring the matrix elements between occupied and unoccupied Eigen states. The imaginary part of the dielectric function due to direct intraband transition can be derived from the Fermi golden rule[35]:

$$\varepsilon_2(\omega) = \frac{4\pi^2 e^2}{\Omega} \lim_{q \to 0} \frac{1}{q^2} \sum_{c,v,k} 2w_k \delta(\varepsilon_{ck} - \varepsilon_{vk} - \omega) \times \langle u_{ck} + e_{\alpha q} | u_{vk} \rangle \langle u_{ck} + e_{\beta q} | u_{vk} \rangle^*$$

(2)

where the indices c and v refer to conduction and valence band states respectively, and $\mu_{ck}$ is the cell periodic part of the wavefunctions at the given k-point. The dipole transition matrix element was calculated from the self-consistent band structure within the PAW[26,27] method.

## Results and discussion

Fig. 1 shows the structure of the single-layer $MoS_2$, and the visualization of the electron localization function (ELF)[36] that gives us insight into the nature of the bonding. The relaxed lattice constant from GGA-PBE method (a=3.178 Å) is very close to the experimental data[37,38] and theoretical value of other authors[4-6]. The S-Mo-S angle is 80.9 degree, which is almost the same as the value of 81.1 degree from Tianshu Li[5] and close to that of 80.69 degree from C. Ataca[39]. The grade of the ELF is encoded using a colour scheme in which high values correspond to red and low values to blue. The site of the Mo cation is characterized by low charge densities, which is in reasonable expectation from pseudopotential calculations. S anions are identified by large red annular regions. This electron localization around S anions indicates an ionic kind of bonding. However, a degree of covalence is present, because the region of high ELF around S anions is not spherically symmetric and exhibits lobes directed toward the Mo ion. Besides, there are also lobes between the adjacent S cations either in the same layer or two layers. So, the bonding between Mo and S is a mixture of ionic and covalent bonds, and there are also slight covalent characteristic bonds between the adjacent S anions.

The electronic structure of single-layer $MoS_2$ was intensively investigated recently by both experiments and theory[1,2,4-8,12,14,38-41]. Andrea Splendiani investigated the electronic and optical properties of single-layer $MoS_2$ structure through optical reflection, Raman scattering, and photoluminescence spectroscopy. A direct bandgap of about 1.75 eV was obtained from the measurements[2]. Kin Fai Mak also revealed a direct gap of 1.90 eV by using the combination of three complementary techniques, with samples prepared both on solid surfaces and as freestanding films[1]. The theoretical bandgaps calculated through density functional theory also slightly differ for different functional. As can be seen from Table 1, Tianshu Li[5] using PBE functional and Trouiller-Martins type norm-conserving pseudopotentials, obtained a bandgap of about 1.78 eV for single-layer $MoS_2$. Emilio Scalise[6] reported a value of 1.72 eV from GGA-PBE method. C. Ataca[3,39] calculated a direct bandgap of 1.63 eV by GGA-PBE method taking into consideration of the van der Waals corrections (vdW)[42], and 1.87 eV by LDA method. Tawinan and Walter[43] employed quasiparticle self-consistent GW method and got a direct bandgap of 2.76 eV.

Table 1. The calculated and experimental bandgap values of the single-

layer MoS$_2$.

| Methods | Exp. | PBE | GGA +vdW | LDA | GW | HSE06 this work | GGA this work |
|---|---|---|---|---|---|---|---|
| Bandgap (eV) | 1.90[1] 1.75[2] | 1.78[5] 1.72[6] 1.58[39] | 1.63[3] | 1.87[3] | 2.76[43] | 2.10 | 1.70 |

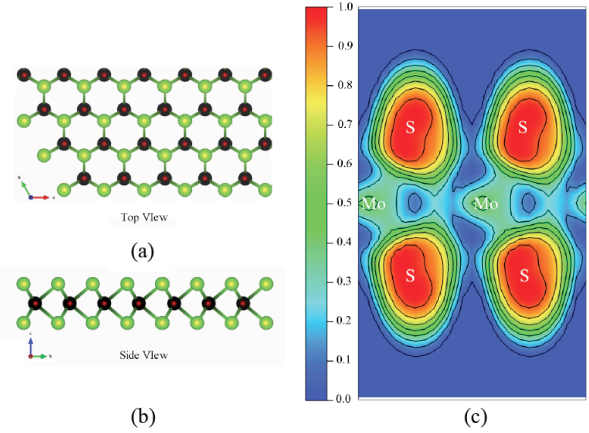

Fig. 1. The structure of single-layer MoS$_2$: a) top view, b) side view; and c) the isosurface of the electron localization function in (110) plane. The contour interval is 0.1.

The experimental researchers have been very careful in order to obtain the freestanding single-layer MoS$_2$, because the electronic structure is sensitive to strain and contact. The dispersion of bandgap in the range of 1.75-1.90 eV in experiments may come from the different preparation and characterization methods, but also suggests the sensitivity of the electronic structure to environments. The theoretical findings on the direct bandgap disperse from 1.63 eV to 2.759 eV for relaxed structures, depending on the exchange and correlation functionals. There is also discrepancy within same methods as can be seen from Table 1. It can be speculated that the electronic structure of the single-layer MoS$_2$ is sensitive to the atomic structure, which is also clarified by the works from Tianshu Li[5] and Emilio Scalise[6] separately. The linear strain coefficient is as much as -112.2±4.2 meV/% under uniaxial tension[5], and the single-layer MoS$_2$ can even become metallic under applied strain.

The calculated direct bandgap is found to be 1.70 eV with the relaxed Mo-S bond length of 2.41 Å in our work, while the direct bandgap from HSE06 method is 2.10 eV with the relaxed Mo-S bond length of 2.39 Å. It is well kwon that the GGA-PBE usually underestimates the bandgap by about 30%[44], while the screened Coulomb potential HSE can describe the bandgap more accurately. However, in the case of MoS$_2$, it seems that the GW method and HSE method as well as the GGA+vdW method failed to reproduce the experimental bandgap, while the GGA-PBE and LDA acquired better performances. It is noted that the dimension of the single-layer MoS$_2$ is one less than bulk, and the interaction between the layers in bulk MoS$_2$ is mainly van der Waals interaction. The LDA method that totally ignores such interactions therefore is more suitable for the single-layer material, as well as the GGA-PBE method. Fig. 2 shows the calculate band structures from GGA-PBE. The VBM (valence band maximum) is lower at Γ point than in K point and there is a direct bandgap at K. We have used GGA-PBE theory level for the following calculations since the HSE and GGA-PBE present the similar characteristics of the band structure for single-layer MoS$_2$.

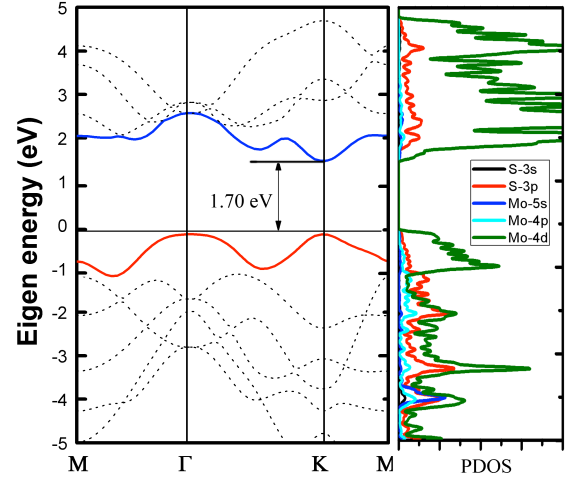

Fig. 2. The band structure and PDOS of the single-layer MoS$_2$ calculated by GGA-PBE method.

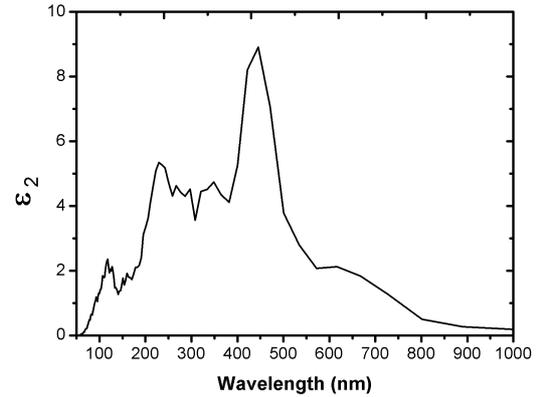

Fig. 3. The imaginary part of the dielectric constant function of the single-layer MoS$_2$.

From the bandgap calculations, it can be seen that the single-layer MoS$_2$ seems to be active for visible light absorption, which can be confirmed by the light adsorption spectrum. Fig. 3 shows the imaginary part of the dielectric constant function of the single-layer MoS$_2$. The imaginary dielectric constant measures the light absorption ability of materials. Since the peaks mostly lie among the visible light range (350 nm-700 nm), the single-layer MoS$_2$ would display high absorption of the sunlight, which is ideal for photocatalysts applications. However, it has a relatively low adsorption rate at the long wavelength end. Such phenomenon originates from the electronic structure of the single-layer MoS$_2$. As can been seen in the PDOS of MoS$_2$ in Fig. 2, the VBM is dominated by Mo-*4d* and marginally from S-*3p*, while the CBM is totally contributed by Mo-*4d*. Whereas, the transition from occupied to unoccupied Mo-*4d* states is symmetrically forbidden. Only the transition from S-*3p* to Mo-*4d* is allowed. So the adsorption peak has a relatively low start from the long wavelength end. The main peak that lies at about 450 nm is speculated to attribute to the transition from Mo-*4d* to S-*3p*. From the dielectric constant of the single-layer $MoS_2$, it can be concluded that this material is suitable for energy harvesting applications as it displays high adsorption in the visible light range.

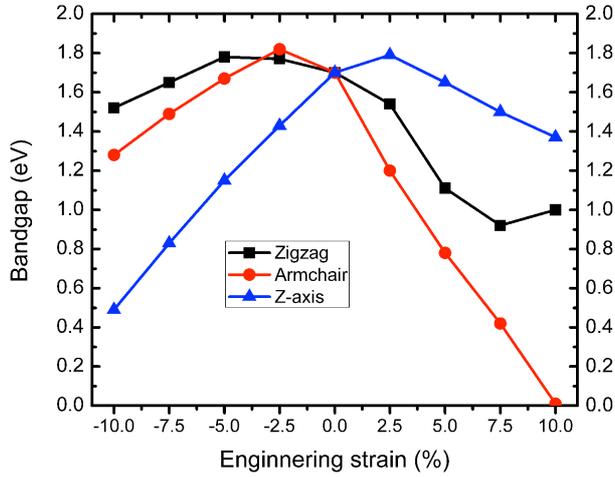

Fig. 4. The evolution of the bandgap of single-layer $MoS_2$ under applied mechanical strain.

The mechanical strain was applied along the zigzag, armchair and z-axis direction, ranging from -10% to 10%, to probe the effects on the electronic structure. The single-layer $MoS_2$ is soft and stretchable, and the strain up to 10% can be achieved without reaching to limits in practice[5]. The evolution of bandgap with respect to the applied strain is shown in Fig.4. The evolution trends of bandgap under tensile strain along armchair direction and zigzag direction are consistent with the findings of Tianshu Li[5] and Emilio Scalise[6]. The bandgap decreases continuously with the increasing strain, and the speed is much faster along armchair direction. It is consistent with the finding of Tianshu Li[5] that the strength along armchair direction is much weaker. The single-layer $MoS_2$ becomes metallic when the strain is big enough as founded by Emilio Scalise[6] and Tianshu Li[5]. Surprisingly, the compressive in-plane strain induced an unusual increase of bandgap, which is followed by a continuous decrease again. Such phenomenon is rarely seen in comparison with the regular variations of bandgap under strain that have been observed in Si[22] and other semiconductors. Furthermore, the evolution trend of bandgap under the strain along z-axis is reversal to that in zigzag direction and armchair direction. The bandgap increases firstly and then decreases with increasing tensile strain, while it decreases steadily under compressive strain.

To reveal the potential application of the single-layer $MoS_2$ as photocatalyst, it is necessary to align the band edge potentials with respect to the redox reaction levels. Since the bandgap of single-layer $MoS_2$ are significantly influenced by mechanical strain. It is important to check the band edge potentials change under strain. It has been shown in Fig. 5 that not only the bandgap changes significantly under strain, but also the band levels of CBM and VBM. First, it is notable that the single-layer $MoS_2$ is a suitable candidate for photocatalyst. The CBM is located at a more negative potential than the proton reduction potential ($H^+/H_2$), while the VBM is positioned more positively than the oxygen reduction potential ($O_2/H_2O$). The bandgap of about 1.7 eV is ideal for visible light active photocatalyst as well as the effective utilization of the solar energy. In a much closer examination, the oxidizing power (denoted as $\Delta_2$ that is defined as the difference between VBM band edge and the oxygen reduction potential, is found to be only 0.07 eV. However, the reducing power (denoted as $\Delta_1$), which is defined as the difference between the proton reduction and the CBM edge potentials is found to be 0.41 eV. The small oxidizing power diminishes and even becomes negative whenever strain is applied in any direction. Thus lose the oxidization ability. While the reducing power gains increase when small in-plane compressive strain or out-of plane tensile strain is applied, which strengthened the reduction ability of the single-layer $MoS_2$. As the main purpose of the photocatalytic reaction is to obtain hydrogen gas, it can be speculated that the appropriate mechanical strain can improve the hydrogen production. However, the imbalance between the reduction reaction and oxidation reaction will prevent the further process of redox reactions. It can only be continued when excessive oxidization species are reduced, which requires the introduction of co-catalyst or electrode.

To enable the single-layer $MoS_2$ as a self-consistent photocatalyst, the reduction reaction and the oxidization reaction should proceed simultaneously in a relatively synchronized pace. Actually, the rate constant of any chemical reaction can be given by Arrhenius equation[45] such as:

$$k = Ae^{-E_a/RT} \qquad (3)$$

where *A* is the pre-exponential factor or simply the prefactor and *R* is the universal gas constant, and $E_a$ corresponds to the reaction energy barrier. The reaction barrier in redox reactions of water mediated by photocatalysts is a complex factor. However, it still can be speculated that the difference between the reaction rate of reduction reaction and oxidation reaction can be minimized by reducing the difference between $\Delta_1$ and $\Delta_2$, and improved efficiency of redox reaction of water mediated by photocatalysts can be achieved. In this context, the single-layer $MoS_2$ is a good candidate for reduction but not oxidation. The big difference between $\Delta_1$ and $\Delta_2$ imbalances the reduction and oxidation reactions, which in turn inhibits the progress of the overall redox reaction. Reducing the difference between $\Delta_1$ and $\Delta_2$ may allow the single-layer $MoS_2$ to be self-consistent photocatalyst and further improve its efficiency for water splitting.

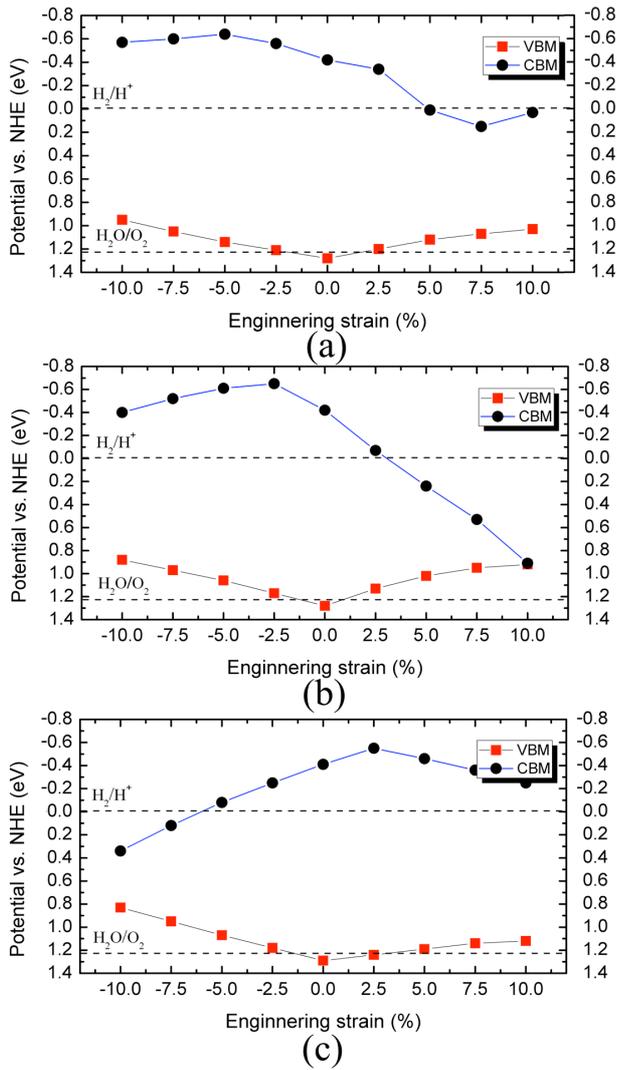

Fig. 5. The evolution of the VBM and CBM under applied strain: a) zigzag direction, b) armchair direction, c) z-axis direction.

If the oxidation power can be promoted, even at a little expanse of reduction power, the overall photocatalytic reaction can be improved. The controlled bandgap engineering is necessary in order to enhance the catalysing efficiency of photocatalysts, and the extrinsic doping of a foreign element into a semiconductor is a promising strategy for modification of band structure. The sensitive dependence of the semiconductor's electronic and optical properties on dopants has provided a powerful approach to explore and apply this material. Besides that, the doping process usually improves the electrical conductivity of semiconductors[9], which is favourable for the reduction of recombination of electrons and holes in the photocatalysts. Our expectation is to shift downward the VBM level of the single-layer $MoS_2$ while slightly lowering or stabilizing the CBM. It needs to create "broken bonds" (holes) in the lattice that are free to move, which is p-type doping. Our idea is to dope the Group V element, phosphorus, which is commonly used as a dopant for semiconductor. The phosphorus has one less electron than sulphur, so it is expected that the bond between the dopant and Mo would be likely weaker than the bond between Mo and S. Thus the holes can be introduced successfully by p-type doping.

Thus, we have added one phosphorus atom in a 4×4×1 supercell with 32 S atoms and 16 Mo atoms, which corresponds to a low concentration of 1.21 wt.%. The structure was fully relaxed, and the band structure as well as the band edge potentials with respect to the redox levels is shown in Fig. 6. After doping, the lowest peak of $3s$ orbital of sulphur was shifted downwards only by 0.04 eV. In this work, the alignment of the doped single-layer $MoS_2$ was done in reference to the lowest peak of $3s$ orbital of the pristine single-layer $MoS_2$. The results in Fig. 6 show successful p-type doping. The bandgap was slightly reduced to 1.62 eV. The $\Delta_1$ was decreased to 0.20 eV, and $\Delta_2$ was increased to 0.19 eV, so that they became very close. The equalization of the reduction power and oxidation power can minimize the difference between the reduction and oxidation reaction, and further improve the catalyzing ability.

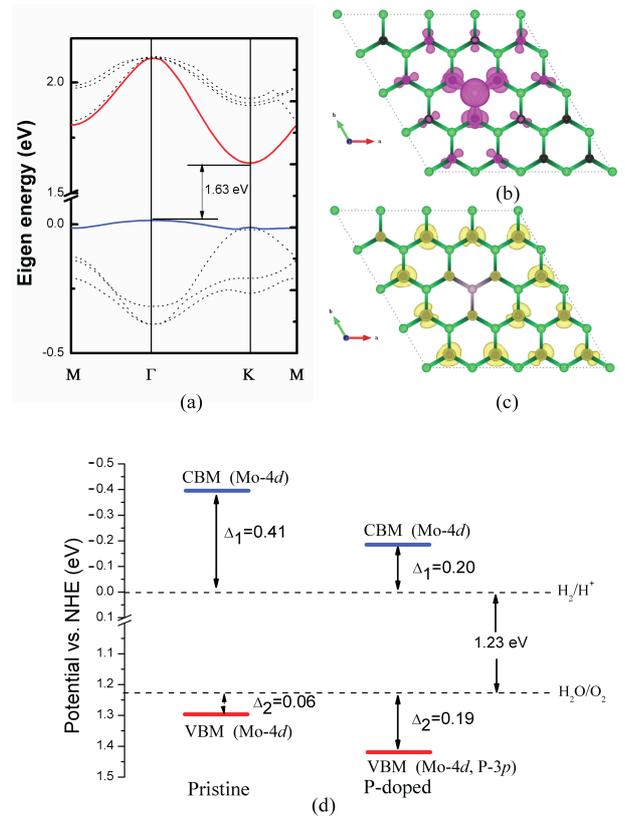

Fig. 6. a) The band structure of phosphorus-doped single-layer $MoS_2$; b) and c) the partial charge density distributions of VBM and CBM of the phosphorus-doped single-layer $MoS_2$, respectively. The black atom is Mo, green is S, pink is P; d) the energy alignment of the phosphorus-doped single-layer $MoS_2$ compared to the pristine.

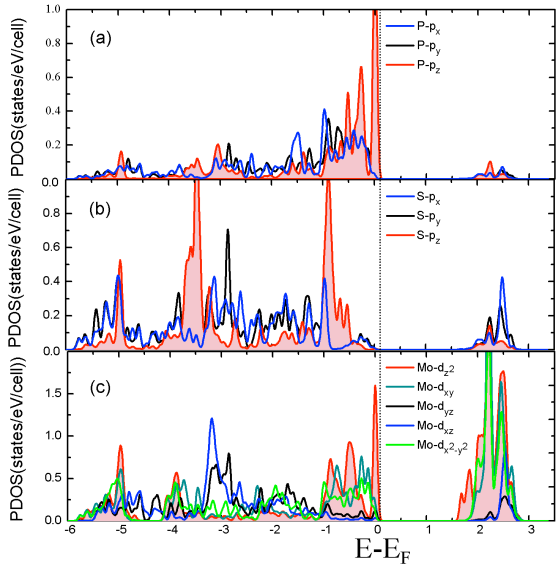

Fig. 7. The PDOS of the phosphorus-doped single-layer $MoS_2$: a) PDOS of P, b) PDOS of S, c) PDOS of Mo.

It can be seen in Fig. 7 that the $p_z$ orbital of P and $d_z^2$ orbital of Mo strongly hybridize just below Fermi level. Compared with S atom, the P atom needs one more electron from Mo to fulfil its $p$ orbitals. So, one empty orbital was created in the $d$ bands of Mo, which corresponds to the state of CBM in the doped system. While P atom accepted the "excess" electron from the Mo atom, which corresponds to state of the VBM. Therefore, the energy level of VBM in $MoS_2$ was lowered by a hole generated from the Mo atom by P doping. After doping, the VBM is dominated by P-$3p$ orbitals and the CBM is governed by Mo-$4d$ orbitals as can be seen from band structure in Fig.6 and PDOS in Fig.7. Now, the transition directly from the VBM to CBM is allowed by parity after doping. Therefore, the light absorption efficiency of the single-layer $MoS_2$ is strengthened in the long wavelength end. What is more, the reduction and oxidation of water will react in the Mo sites and S sites separately after p-type doping. As can be seen in Fig.6 b and c, the electron of VBM (red isosurface) mostly lies around P atom and adjacent Mo atoms, while the electron of CBM (yellow isosurface) gathers around the Mo atoms that are far away from P atoms. Such charge density distribution can not only separate the reduction and oxidation sites during water splitting, but also depress the recombination of electrons and holes. So the catalysing efficiency would be greatly improved by p-type doping.

## Conclusions

In summary, the potential application of the single-layer $MoS_2$ as photocatalyst was revealed in this work. The pristine single-layer $MoS_2$ can be a good photocatalyst but needs the introduction of co-catalyst or electrode. The applied small in-plane compressive strain or out-of plane tensile strain can improve its hydrogen production ability. While the p-type doping can significantly improve the catalyzing ability by allowing the direct electron transition from VBM to CBM and by separating the reduction and oxidations sites. Furthermore, the difference of energy barrier between the reduction reaction and oxidation reaction can be minimized by equalizing $\Delta_1$ and $\Delta_2$ through p-type doping, which improves the overall photocatalytic redox reaction. The light absorption was also enhanced in long wavelength end of visible light with p-type doping. Thus facilitates the single-layer $MoS_2$ as an efficient self-consistent visible light photocatalyst.

## Acknowledgements

We would like to acknowledge the Swedish Energy Agency, Swedish Research Council (VR) and Stiffelsen J. Gust Richerts Minne (SWECO) for financial support. Yan-ling Li is also thankful to National Science Foundation of China (Grant no. 11047013), the Priority Academic Program Development of Jiangsu Higher Education Institutions (PAPD), and Jiangsu Overseas Research & Training Program for University Prominent Young & Middle-aged Teachers and Presidents. Yunguo Li also thanks to Chinese Scholarship Council (CSC). SNIC and UPPMAX are acknowledged for providing computing time.

[a] Applied Materials Physics, Department of Material science and Engineering, Royal Institute of Technology (KTH), S-100 44 Stockholm, Sweden; E-mail: Yunguo@kth.se
[b] School of Physics and Electronic Engineering, Jiangsu Normal University, 221116, Xuzhou, People's Republic of China;
[c] Condensed Matter Theory Group, Department of Physics and Astronomy, Box 516, Uppsala University, 751 20 Uppsala, Sweden;